\documentclass[fleqn]{elsart}

\usepackage{graphicx} 
\usepackage{booktabs, multirow}
\usepackage{epsfig,amsmath,amsfonts}
\usepackage{dcolumn}  
\usepackage{threeparttable} 
\usepackage{bm}
\usepackage{ifpdf}
\usepackage{graphicx} 
\usepackage{xspace}
\usepackage{amssymb}
\journal{Phys. Lett. B}

\newcommand{\ks}{K^0_S}
\newcommand{\ee}{e^+e^-}
\newcommand{\ggr}{\gamma\gamma\rightarrow}
\newcommand{\GeV}{{\rm GeV}}
\newcommand{\GeVc}{{\rm GeV/}\ensuremath{c}}
\newcommand{\costs}{\cos\theta^*}
\ifpdf
\usepackage[%
  pdftitle={A study of $\ggr\ks\ks$ Production at energies of 2.4 - 4.0 GeV at Belle},%
  pdfauthor={W.T.~Chen},%
  pdfsubject={A study of $\ggr\ks\ks$ Production at energies of 2.4 - 4.0 GeV at Belle},%
  pdfkeywords={Two-photon collisions; Mesons; QCD; Charmonium},%
  pdfstartview=FitH,%
  bookmarks=true,%
  bookmarksopen=true,%
  breaklinks=true,%
  colorlinks=true,%
  linkcolor=blue,anchorcolor=blue,%
  citecolor=blue,filecolor=blue,%
  menucolor=blue,pagecolor=blue,%
  urlcolor=blue]{hyperref}
\else
\usepackage[%
  breaklinks=true,%
  colorlinks=true,%
  linkcolor=blue,anchorcolor=blue,%
  citecolor=blue,filecolor=blue,%
  menucolor=blue,pagecolor=blue,%
  urlcolor=blue]{hyperref}
\fi

\makeatletter
\def\elsartstyle{%
    \def\normalsize{\@setfontsize\normalsize\@xiipt{14.5}}
    \def\small{\@setfontsize\small\@xipt{13.6}}
    \let\footnotesize=\small
    \def\large{\@setfontsize\large\@xivpt{18}}
    \def\Large{\@setfontsize\Large\@xviipt{22}}
    \skip\@mpfootins = 18\p@ \@plus 2\p@
    \normalsize
}
\@ifundefined{square}{}{\let\Box\square}
\makeatother

\begin{document}

\begin{frontmatter}
\vspace*{-3\baselineskip}
\begin{flushleft}
 \resizebox{!}{3cm}{\includegraphics{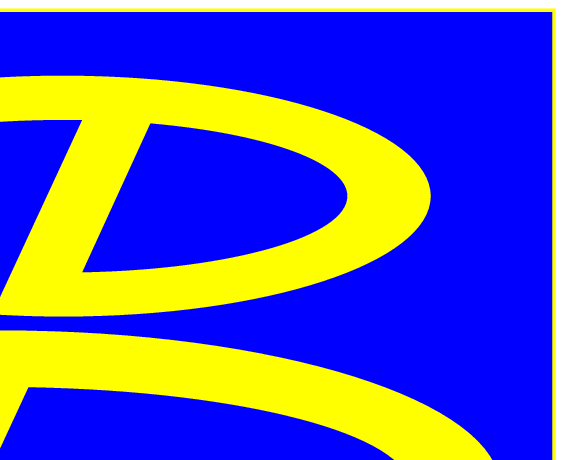}}
\end{flushleft}
\vspace*{-3cm}
\begin{flushright}
BELLE-CONF-0660 \\
Belle Prerpint 2007-20 \\
KEK   Preprint 2007-7 \\
hep-ex/0609042 \\
\end{flushright}
\vspace*{2cm}

\title{A study of $\ggr\ks\ks$ production at energies from 2.4 to 4.0 GeV at Belle}


\collab{Belle Collaboration}
  \author[NCU]{W.~T.~Chen}, 
  \author[KEK]{K.~Abe}, 
  \author[TohokuGakuin]{K.~Abe}, 
  \author[KEK]{I.~Adachi}, 
  \author[Tokyo]{H.~Aihara}, 
  \author[BINP]{D.~Anipko}, 
  \author[BINP]{V.~Aulchenko}, 
  \author[Sydney]{A.~M.~Bakich}, 
  \author[Melbourne]{E.~Barberio}, 
  \author[Lausanne]{A.~Bay}, 
% ...
and

\address[BINP]{Budker Institute of Nuclear Physics, Novosibirsk, Russia}
% ...
\address[KEK]{High Energy Accelerator Research Organization (KEK), Tsukuba, Japan}
% ...
\address[Lausanne]{Swiss Federal Institute of Technology of Lausanne, EPFL, Lausanne, Switzerland}
% ...
\address[Melbourne]{University of Melbourne, Victoria, Australia}
% ...
\address[NCU]{National Central University, Chung-li, Taiwan}
% ...
\address[Sydney]{University of Sydney, Sydney, NSW, Australia}
% ...
\address[TohokuGakuin]{Tohoku Gakuin University, Tagajo, Japan}
% ...
\address[Tokyo]{Department of Physics, University of Tokyo, Tokyo, Japan}
% ...

\begin{abstract}
$K^0_SK^0_S$ production in two-photon collisions has been studied
using a 397.6~fb$^{-1}$ data sample
collected with the Belle detector at the KEKB $e^+e^-$ collider.
For the first time the cross sections are measured in the two-photon 
center-of-mass energy range between 2.4~$\GeV$ and $4.0~$\GeV~~and 
angular range $|\cos\theta^*|<0.6$.
Combining the results with measurements of $\ggr K^+K^-$ from Belle,
we observe that the cross section ratio $\sigma(\ks\ks)/\sigma(K^+K^-)$ 
decreases from $\sim$0.13 to $\sim$0.01 with increasing energy.
Signals for the $\chi_{c0}$ and $\chi_{c2}$ charmonium states are also 
observed.
\end{abstract}

\end{frontmatter}


{\renewcommand{\thefootnote}{\fnsymbol{footnote}}}
\section{Introduction}
Exclusive processes with hadronic final states in two-photon collision are an excellent probe 
to test various model calculations 
motivated by perturbative and non-perturbative QCD.
As shown by Brodsky and Lepage (BL) \cite{BL}, at sufficiently
large two-photon center-of-mass energy $\sqrt{s}$ and momentum transfer from the initial photon to the produced meson $t$,
the leading term of the amplitude for the process $\ggr
M\overline{M}$, where $M$ denotes a meson, 
can be expressed as a hard scattering amplitude for 
$\ggr q\bar{q}q\bar{q}$ times the leading term meson electromagnetic 
form factor.
For mesons with zero helicity their calculation 
gives the following dependence on $s$ 
and scattering angle $\theta^*$: 
\begin{eqnarray}\label{eq:BL1}
\frac{d\sigma^{\rm lead}}{d|\cos\theta^*|}=
16\pi\alpha^2\frac{|F^{\rm lead}_M(s)|^2}{s}\times\hspace{6cm} \nonumber \\
\Big\{\frac{[(e_1-e_2)^2]^2}{(1-\cos^2\theta^*)^2}+ 
\frac{2(e_1e_2)[(e_1-e_2)^2]}{1-\cos^2\theta^*}g(\theta^*)+
2(e_1e_2)^2g^2(\theta^*)\Big\} ,
\end{eqnarray}
where $e_1$ and $e_2$ are the quark charges 
(i.e., mesons have charges $\pm(e_1-e_2)$), and explicit forms
of the leading term meson form factor 
$F^{\rm lead}_M(s)$ $(F^{\rm lead}_M(s) \sim 1/s~{\rm at}~s \to \infty)$ 
and the function $g(\theta^*)$ can be found in Refs.~\cite{BL,BC}. 
Eq. (\ref{eq:BL1}) implies that the angular distribution of neutral meson pairs, 
unlike that for charged meson pairs which is dominated by $\sim\sin^{-4}\theta^*$ terms, 
is directly determined by the shape of $g(\theta^*)$ and the value of $F^{\rm lead}_M(s)$.
Later, Benayoun and Chernyak (BC) \cite{BC} used a factorization hypothesis
similar to the BL calculation 
but further improved the treatment of the effects of SU(3) symmetry breaking;
their predictions appeared to be in good agreement with the subsequent 
measurements of
$\ggr\pi^+\pi^-$ and $\ggr K^+K^-$~\cite{aleph,nkzw}.

Recently, Diehl, Kroll and Vogt (DKV) \cite{DKV} considered the 
consequences of the assumption that at  intermediate energies the 
amplitudes for the process $\ggr M\overline{M}$ are dominated by so-called 
handbag contributions. The handbag amplitude is expressed as the product of 
an amplitude for the hard $\ggr q\bar{q}$ 
subprocess times an unknown form factor $R_{M\overline{M}}(s)$ describing 
the soft transition from the $q\bar{q}$ to the meson pair.
In~\cite{DKV} the differential cross section is given by
\begin{equation} \label{eq:DKV1}
\frac{d\sigma}{d|\cos\theta^*|}(\gamma\gamma\rightarrow M\overline{M})=
\frac{8\pi\alpha^2}{s}\frac{1}{\sin^4\theta^*}|R_{M\overline{M}}(s)|^2 ,
\end{equation}
where the meson annihilation form factor $R_{M\overline{M}}(s)$ is
not calculated in Ref.~\cite{DKV} but is instead obtained by fitting the data;
the magnitude of $R_{M\overline{M}}(s)$ for different mesons can be linked by using SU(3) and isospin symmetry.
The validity of this approach has recently been criticized in 
Ref.~\cite{chernyak}.

Earlier, the Belle Collaboration performed a high-statistics measurement
of the cross sections for the processes $\gamma \gamma \to \pi^+\pi^-$
and  $\gamma \gamma \to K^+K^-$~\cite{nkzw} in the $W$(=$\sqrt{s}$) range 2.4 GeV~$<~W~<$~4.1 GeV.
Analysis of the data showed that in this $W$ range
the $W$ dependence of the cross section
is consistent with that predicted by the leading term 
QCD calculations~\cite{BL,BC}.  
Here we report a measurement of the cross section for $\ggr\ks\ks$ at
$2.4~\GeV < W < 4.0~\GeV$ and $|\cos \theta^*| < 0.6$ with a data sample 
of 397.6~fb$^{-1}$ collected at or near the $\Upsilon(4S)$ resonance,
accumulated with the Belle detector \cite{Belle_D1}  
at the KEKB asymmetric-energy $e^+e^-$ collider~\cite{Belle_D2}.
This measurement can provide important information
that complements previous studies and sheds light on 
how the two-photon mass and angular
distributions of such cross sections depend on the flavor of the produced mesons.

The Belle detector is a large-solid-angle magnetic spectrometer. 
Momenta of charged tracks are measured with a central drift 
chamber (CDC), located in a uniform 1.5 T magnetic field
which surrounds the interaction point (IP) and subtends the polar 
angle range $17^{\circ}<\theta_{\rm{lab}}<150^{\circ}$, 
where $\theta_{\rm{lab}}$ is a scattering angle in the laboratory frame.
The trajectories of the charged tracks near the interaction point 
are provided by the CDC and the silicon vertex detector (SVD).
Energy measurement of electromagnetically interacting 
particles is performed in an electromagnetic calorimeter (ECL) 
made up of CsI(Tl) crystals. The detector is described in detail
elsewhere~\cite{Belle_D1}.
\section{Event Selection}
Exclusive $\ks\ks$ pairs
are produced in quasi-real two-photon collisions
through the process $\ee\rightarrow\ee\ggr\ee\ks\ks$, 
where the scattered $e^+$ and $e^-$ are lost down the beampipe, 
and only the two $\ks$ mesons are detected. 

We select $\ggr\ks\ks$ candidate events in two stages. At stage I the 
following requirements are applied:
\begin{itemize}
\item
exactly four charged tracks with zero net charge of which at least two 
have $p_t>0.3$ GeV/$c$, $dr<1$ cm, $|dz|<5$ cm, where $p_t$ is the 
transverse momentum in the laboratory frame and $dr$ and $dz$ are the
radial and axial coordinates of the point of closest approach of the track to the nominal IP, 
respectively, and the $z$-axis is the direction opposite
to the positron beam axis;
\item
the sum of the magnitudes of the momenta of all tracks, $\Sigma p$,
and the total energy deposit in the ECL are less than 6~$\GeVc$ and 
6~$\GeV$, respectively;
\item
the invariant mass of these four tracks is less than 4.5~$\GeVc^2$, 
and the missing mass squared of the event is greater than 2~$\GeV^2/c^4$;
\end{itemize}

\begin{figure}[htb]
\begin{center}
\includegraphics[width=0.5\textwidth]{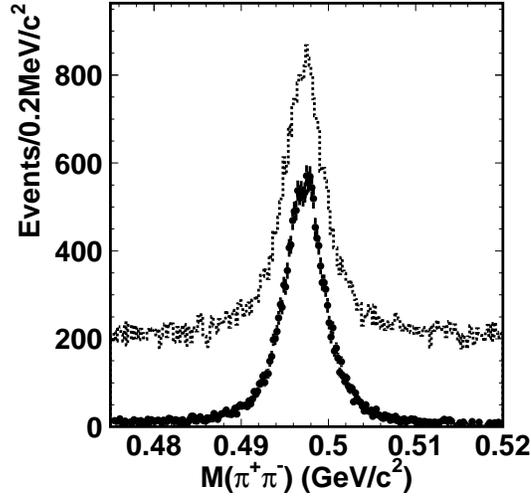}
\caption{The $\pi^+\pi^-$ invariant mass spectrum for $\ks$ candidates 
after stage I (dotted histogram) and stage II (points with error bars) $\ks$ selection. 
Here events are selected in the range $W$=2.3-4.5~GeV and $|\Sigma\mathbf{p}^{ee}_t|<0.25$~GeV/$c$,
where $W$ and $|\Sigma\mathbf{p}^{ee}_t|$ are calculated by assuming all tracks are charged pions. }
\label{fig:ks}
\end{center}
\end{figure}

\begin{figure}[htb]
\begin{center}
\includegraphics[width=0.8\textwidth]{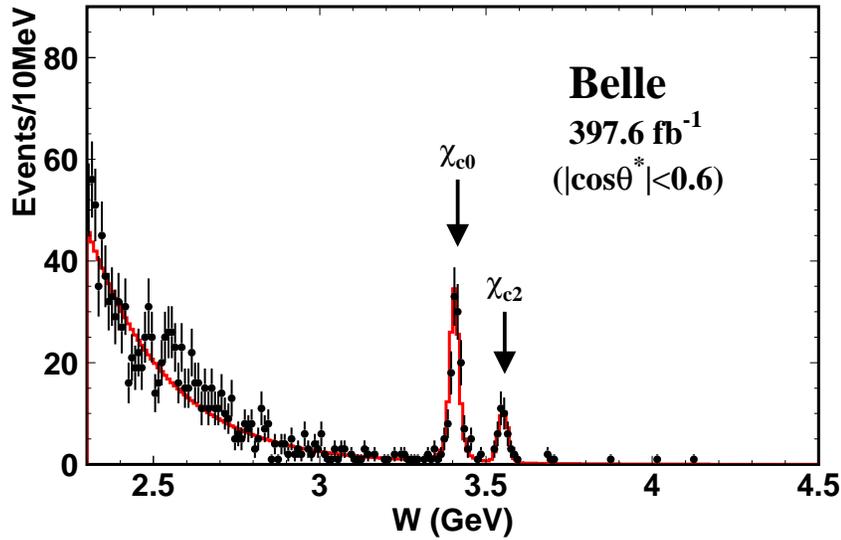}
\caption{$\ks\ks$ mass spectrum before background subtraction. 
The curves show the fit result described later in the section on 
$\chi_{cJ}$ resonances.}
\label{fig:ksks}
\end{center}
\end{figure}

At stage II pairs of oppositely charged tracks without particle 
identification 
are used to reconstruct $K^0_S\rightarrow \pi^+\pi^-$ decays.
To distinguish $\ggr\ks\ks$ events from other four-track background sources 
such as $\ggr2(\pi^+\pi^-)$, $\ggr2(K^+K^-)$, and $\ggr K^+K^-\pi^+\pi^-$ 
that have no $\ks$ candidates,
two different sets of selections are applied to the $\ks$
candidates with high (low) momentum, i.e. with momentum $\gtrsim$~1.5~$\GeVc$ (0.5-1.5~$\GeVc$):
$dr$ is required to be larger than 0.02 (0.03) cm for both charged tracks;
the $\pi^+\pi^-$ vertex is required to be displaced from the IP by a 
minimum transverse distance of 0.22 (0.08) cm. 
The mismatch in the $z$ direction at the $\ks$ vertex point for the 
$\pi^+\pi^-$ tracks must be less than 2.4 (1.8) cm;
the direction of the pion-pair momentum must also agree with the 
direction from the IP to the  vertex to within 0.03 (0.1) rad. 
To evaluate the background and calculate efficiencies, we use 
a Monte Carlo simulation (MC) of the detector response based on
GEANT3 \cite{GEANT}. The TREPS code~\cite{TREPS} is used for 
$\ggr\ks\ks$ event generation and the background $\ggr 2(\pi^+\pi^-)$, $\ggr 2(K^+K^-)$ event generation. 
From MC simulation, 
with the described $\ks$ selection above the $\ks$ signal efficiency can
reach $\sim$ 80\% while the background is reduced by a factor of $10^5$.
Thus the four-track backgrounds can be eliminated efficiently after
our event selection.
The resolution in the reconstructed $\ks$ mass is 4 MeV/c$^2$,
and only candidates for which $|M(\pi^+\pi^-)-m_{\ks}|<13$ MeV/c$^2$ 
are selected. 
Finally, we require that the sum of the transverse momentum vectors of all tracks
in the c.m. frame of the $\ee$ beams, 
$|\Sigma\mathbf{p}^{ee}_t|$, be smaller than 0.1~$\GeVc$ (momentum balance).
$W$ is calculated from the invariant mass of the $\ks\ks$ pair, 
and $|\cos\theta^*|$ is obtained from the $\ks$ scattering angle with respect to
the incident axis of the electron in the $\gamma\gamma$ c.m. frame, 
which approximates the direction of the incoming photon. 
Figure~\ref{fig:ks} shows the 
$\pi^+ \pi^-$ invariant mass spectra after stage I  
and stage II selection.
After applying the above selections,
we find 981 $\ks\ks$ candidates in the range 2.4~$\GeV$$<W<$~4.0~$\GeV$ 
and $|\costs|<0.6$.
The $W$ distribution is shown in Fig.~\ref{fig:ksks}.
Clear signals for the $\chi_{c0}$ and $\chi_{c2}$ resonances are observed.

\section{Background subtraction}
\begin{figure}[htb]
\begin{center}
\includegraphics[width=0.8\textwidth]{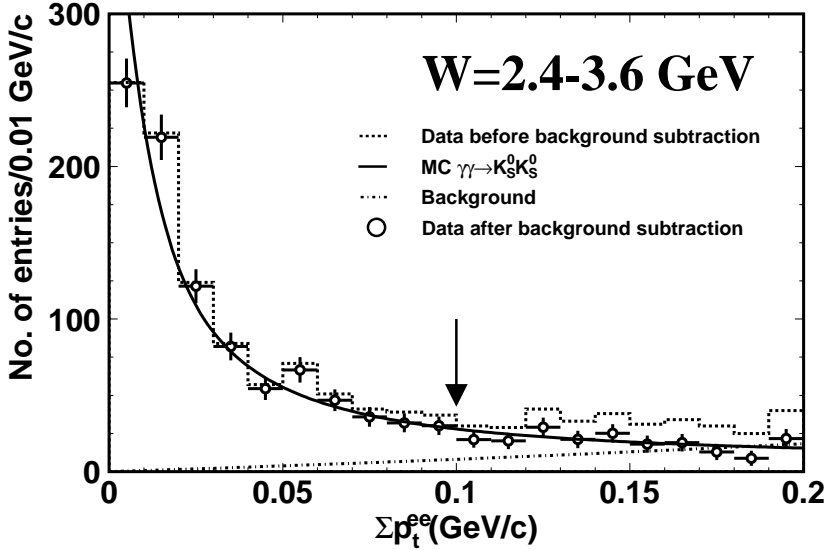}
\caption{$|\Sigma\mathbf{p}^{ee}_t|$ distribution for $\ks\ks$ candidates.
The dotted histogram and points with error bars indicate the 
distribution of events 
before and after background subtraction, respectively.
The dot-dashed line is the background distribution, which is 
obtained from the fit to the difference between MC and data.
The solid curve shows the signal MC
distribution, which is normalized to the number of signal candidates in the three 
leftmost bins.
The arrow indicates the upper boundary of the $|\Sigma\mathbf{p}^{ee}_t|$ requirement 
for the signal.}
\label{fig:nonex}
\end{center}
\end{figure}

The background contamination from events where additional particles 
accompany the two detected $\ks$ mesons
-- so-called non-exclusive backgrounds -- should be also estimated.
Because of the available phase space, such events are expected to have 
a $|\Sigma\mathbf{p}^{ee}_t|$ distribution that is close to zero at 
$|\Sigma\mathbf{p}^{ee}_t|$=0 and increases with $|\Sigma\mathbf{p}^{ee}_t|$.  
This feature is verified in the $\ggr\ks\ks\pi^0$ (which is the dominant background) MC and data sample, 
where the MC sample is generated by using GGLU code \cite{GGLU}.
We assume that the $|\Sigma\mathbf{p}^{ee}_t|$ distribution of the
non-exclusive background can be parameterized by

\[ f(x)= \left\{
\begin{array}{rl}
cx, & x \le 0.05~(\mbox{GeV/}c) \\
ax  + b, & x \ge 0.05~(\mbox{GeV/}c)
\end{array} \right. \] 
constrained by $0.05c=0.05a + b$.
We fit the function $f(x)$ to the difference between data and 
signal MC distributions which is normalized to the data below 0.03 GeV/$c$ 
where the background contribution
is negligibly small (Fig.~\ref{fig:nonex}).
Using the data sample with $|\Sigma\mathbf{p}^{ee}_t|=$0.5-1.0~GeV/$c$ 
we verify that there is no $\theta^*$ dependence of the shape.
Using the fit results,
the estimated background, which is 
$4.1\pm0.1\%$, $3.6\pm0.2\%$, and $2.6\pm0.3\%$ for 
$W$=2.4-2.6~GeV, 2.6-2.8~GeV, 2.8-3.3~GeV, respectively,
is subtracted in each $W$ bin (the errors are statistical only). 
For W=3.6-4.0 GeV, the background is set to zero since
the data sample is too small to apply the procedures described above.

Finally, 952 signal events remain in the signal region 
$|\Sigma\mathbf{p}^{ee}_t|<0.1$~GeV/$c$ after background subtraction.

\section{Cross sections of the process $\ggr\ks\ks$ 
for 2.4~$\GeV~<~W~<~$4.0~$\GeV$}
The differential cross section for two-photon production of the final state 
$X$ in electron-positron collisions is given by
\begin{eqnarray}
\frac{d\sigma}{d|\cos\theta^*|}(W,|\cos\theta^*|;\ggr X)=
\frac{\Delta N(W, |\cos\theta^*|; e^+e^-\rightarrow e^+e^-X)}
{\mathcal{L}_{\gamma\gamma}(W)\Delta W \Delta|\cos\theta^*|
\epsilon(W, |\cos\theta^*|)\mathcal{L}_{\rm{int}}} ,
\end{eqnarray}
where $\Delta N$ and $\epsilon$ denote the number of signal events 
after background subtraction and 
the product of detection and trigger efficiencies, respectively.
The integrated luminosity of this experiment, ${\mathcal L}_{\rm int}$,
is 397.6 fb$^{-1}$ and is determined with a systematic 
uncertainty of 1.4\%.
The luminosity function $\mathcal{L}_{\gamma\gamma}$, as a function of
$W$, is defined by
\begin{equation}
\mathcal{L}_{\gamma\gamma}(W)=\frac{\frac{d\sigma}{dW}
(W;e^+e^-\rightarrow e^+e^-X)}{\sigma(W;\ggr X)} .
\end{equation}

\begin{figure*}[htb]
\begin{center}
\includegraphics[width=0.3\textwidth]{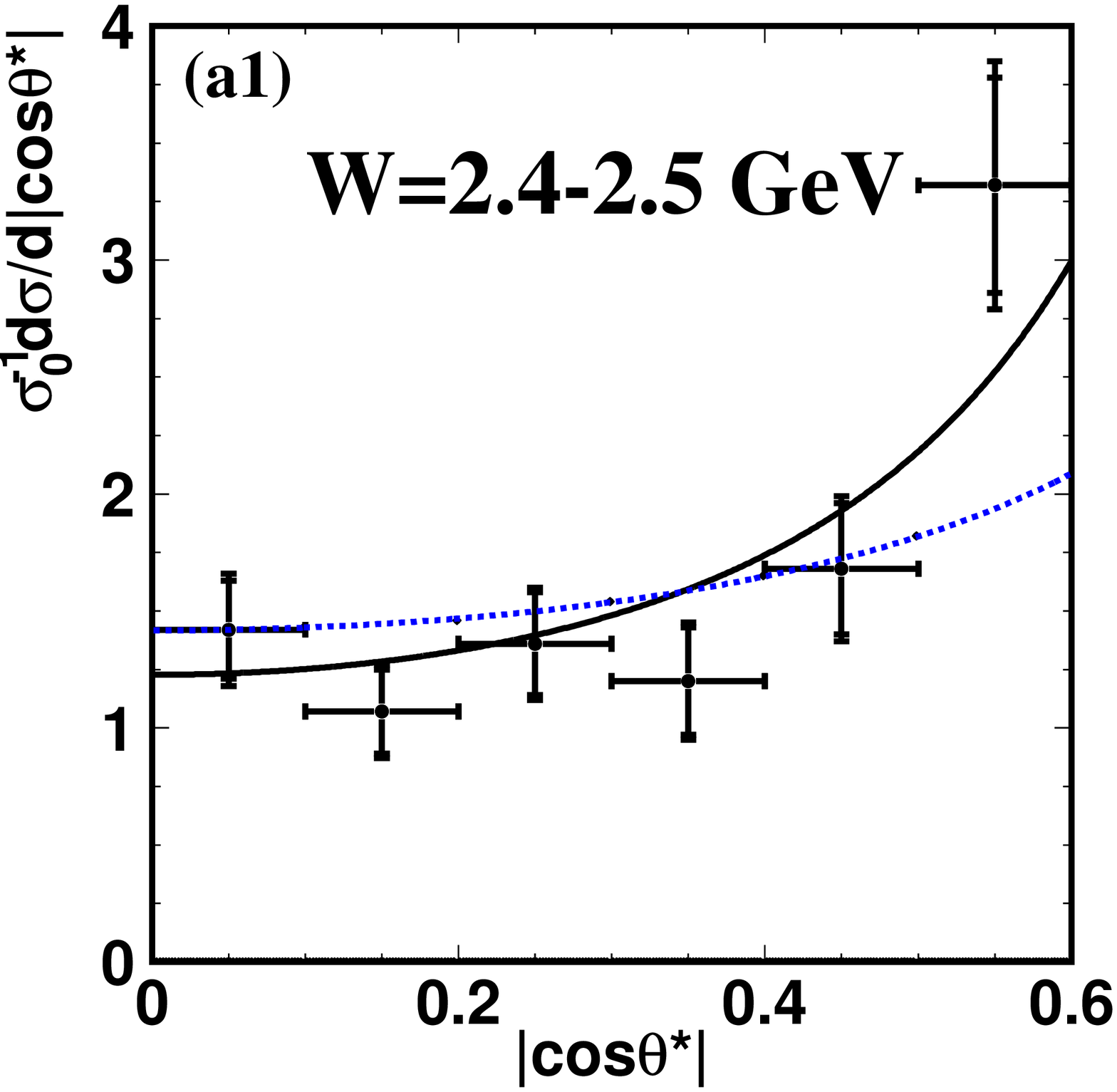}
\includegraphics[width=0.3\textwidth]{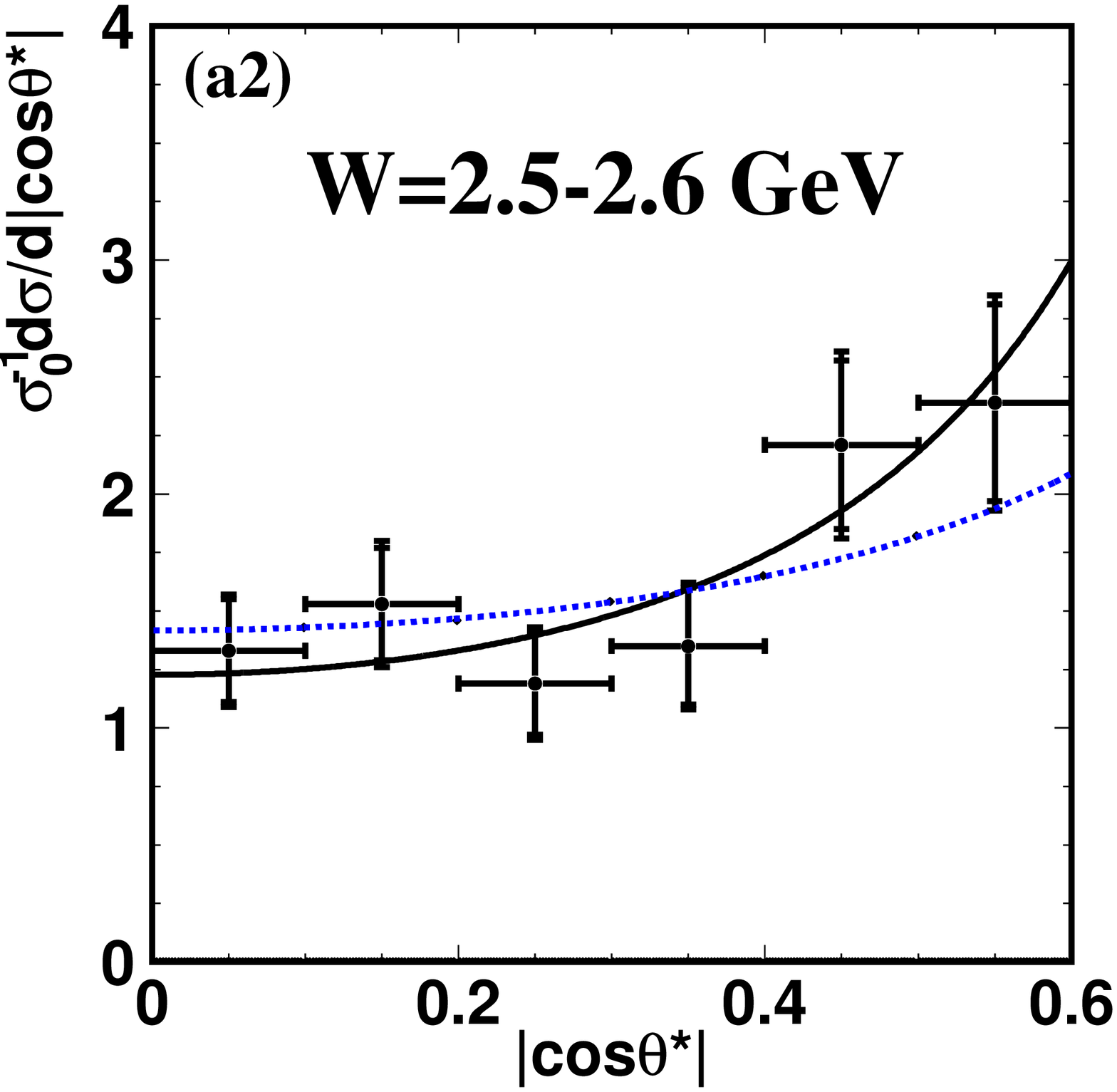}
\includegraphics[width=0.3\textwidth]{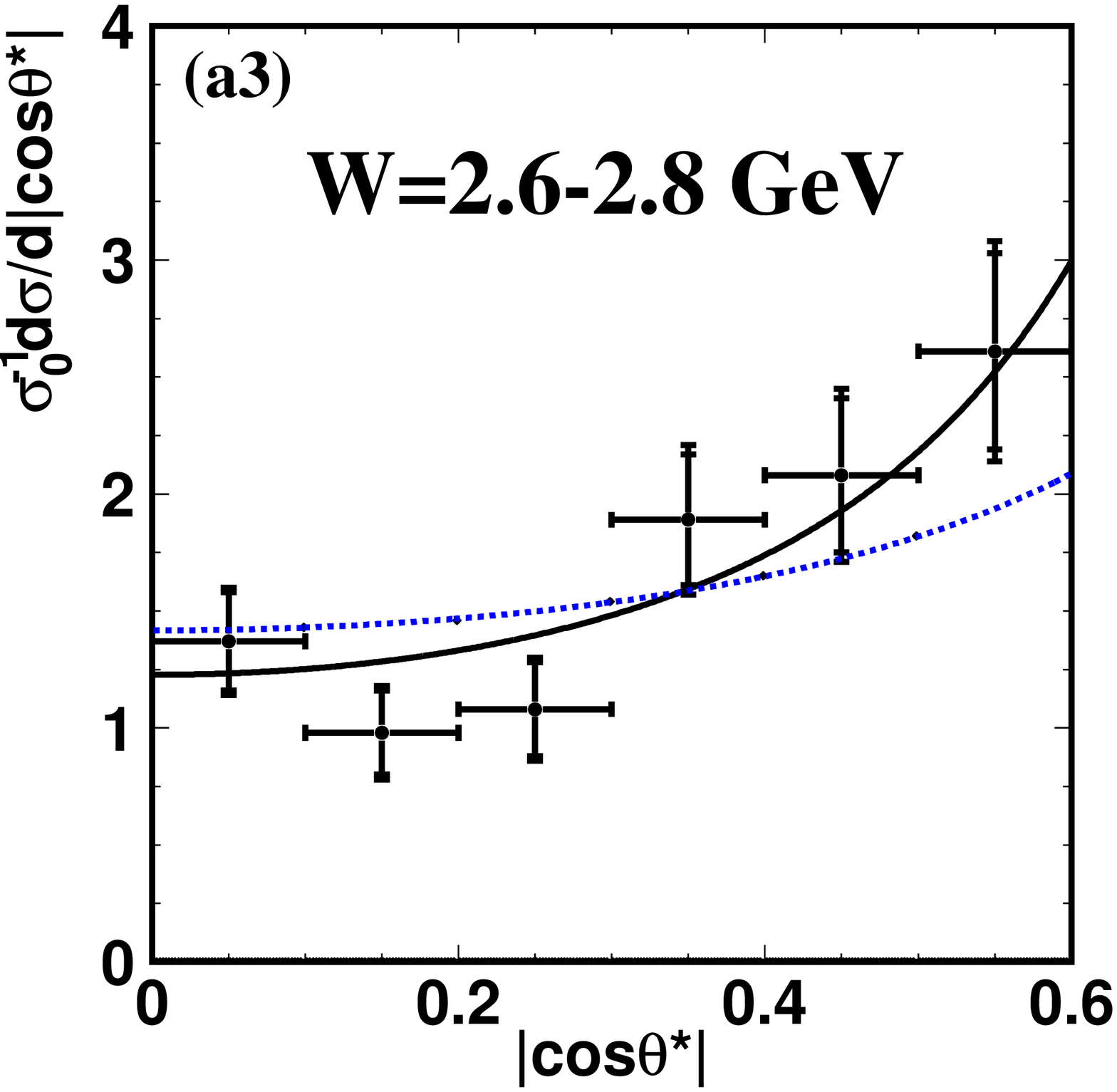} \\
\includegraphics[width=0.3\textwidth]{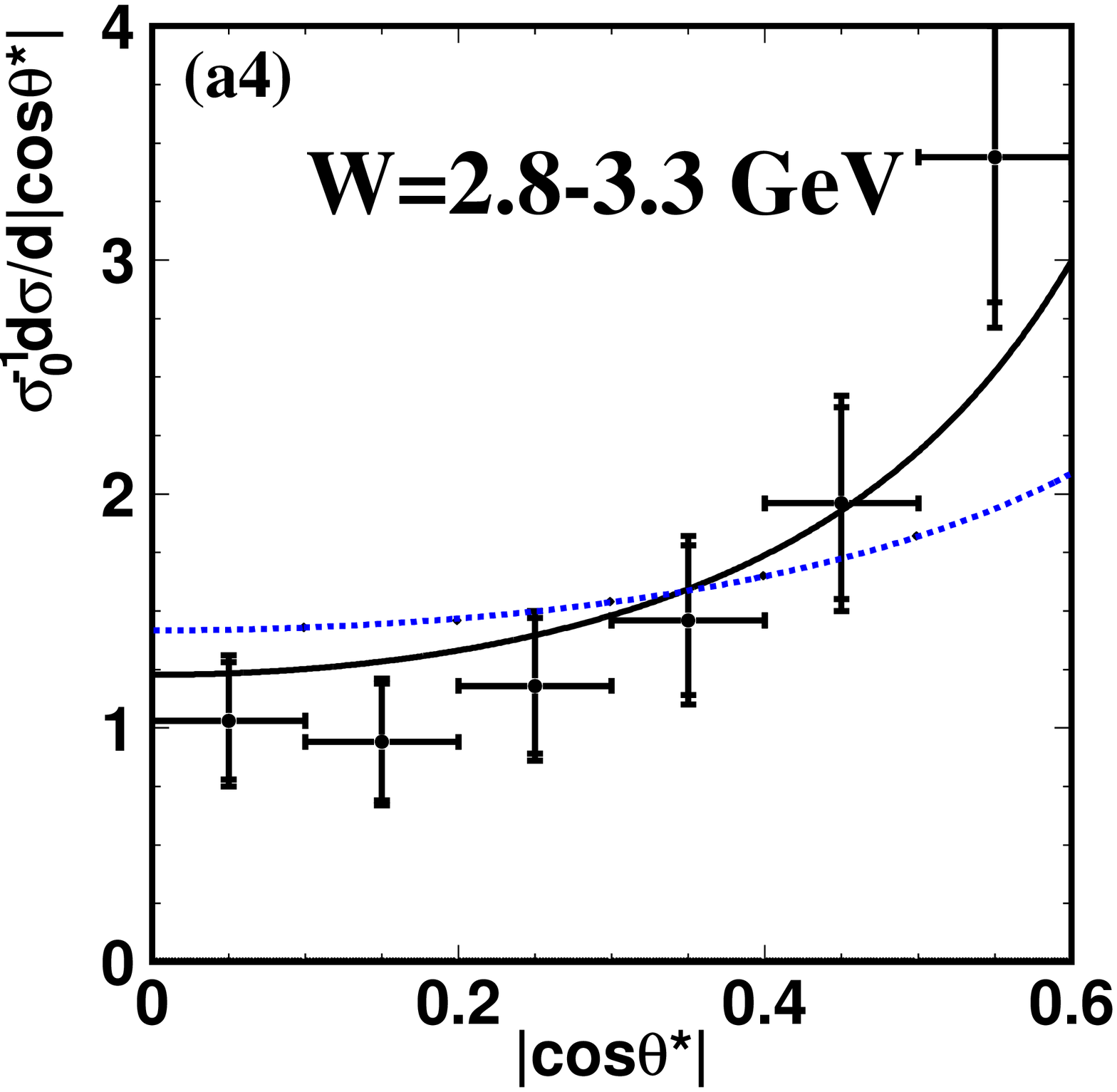} 
\includegraphics[width=0.3\textwidth]{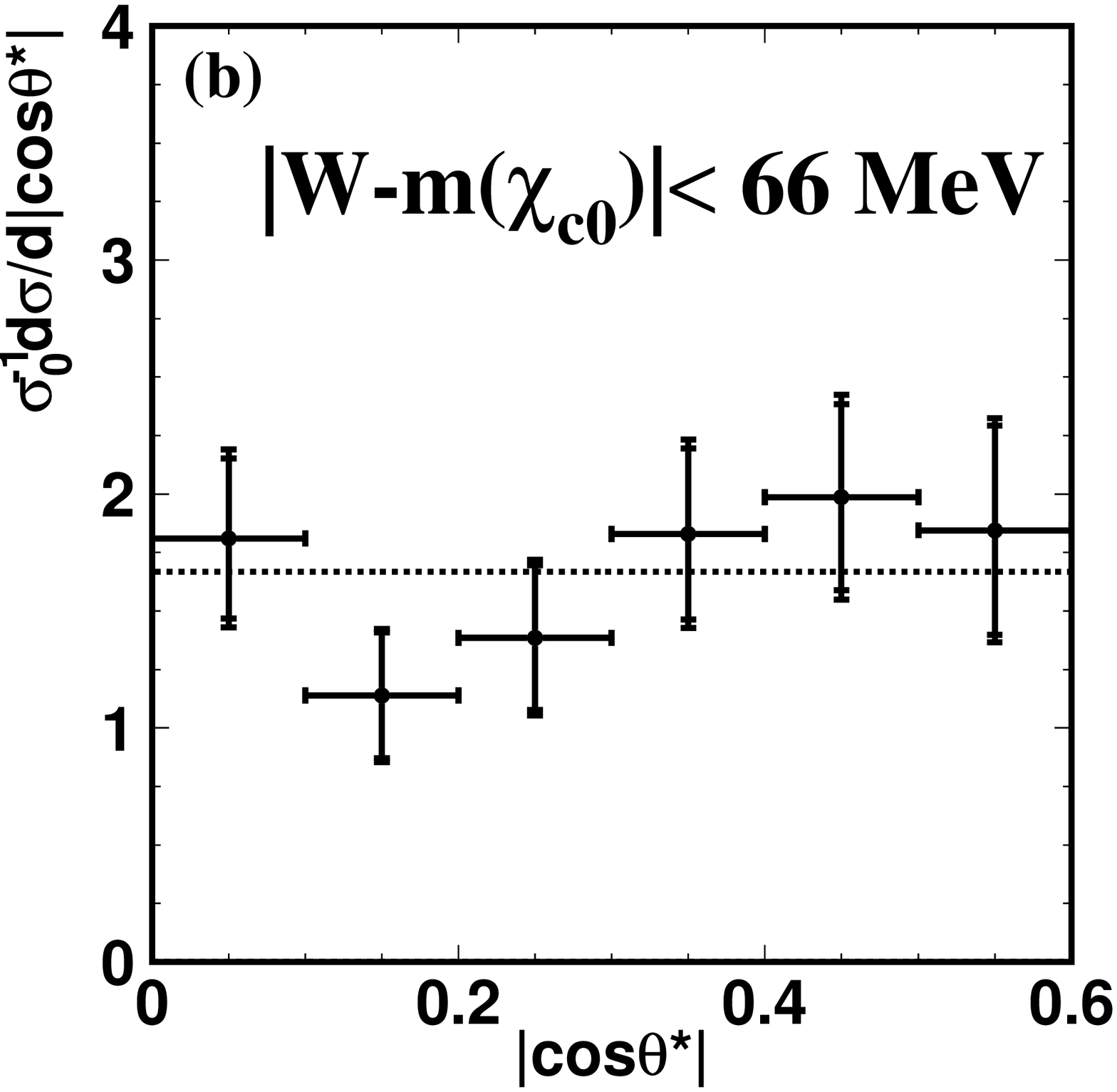}
\includegraphics[width=0.3\textwidth]{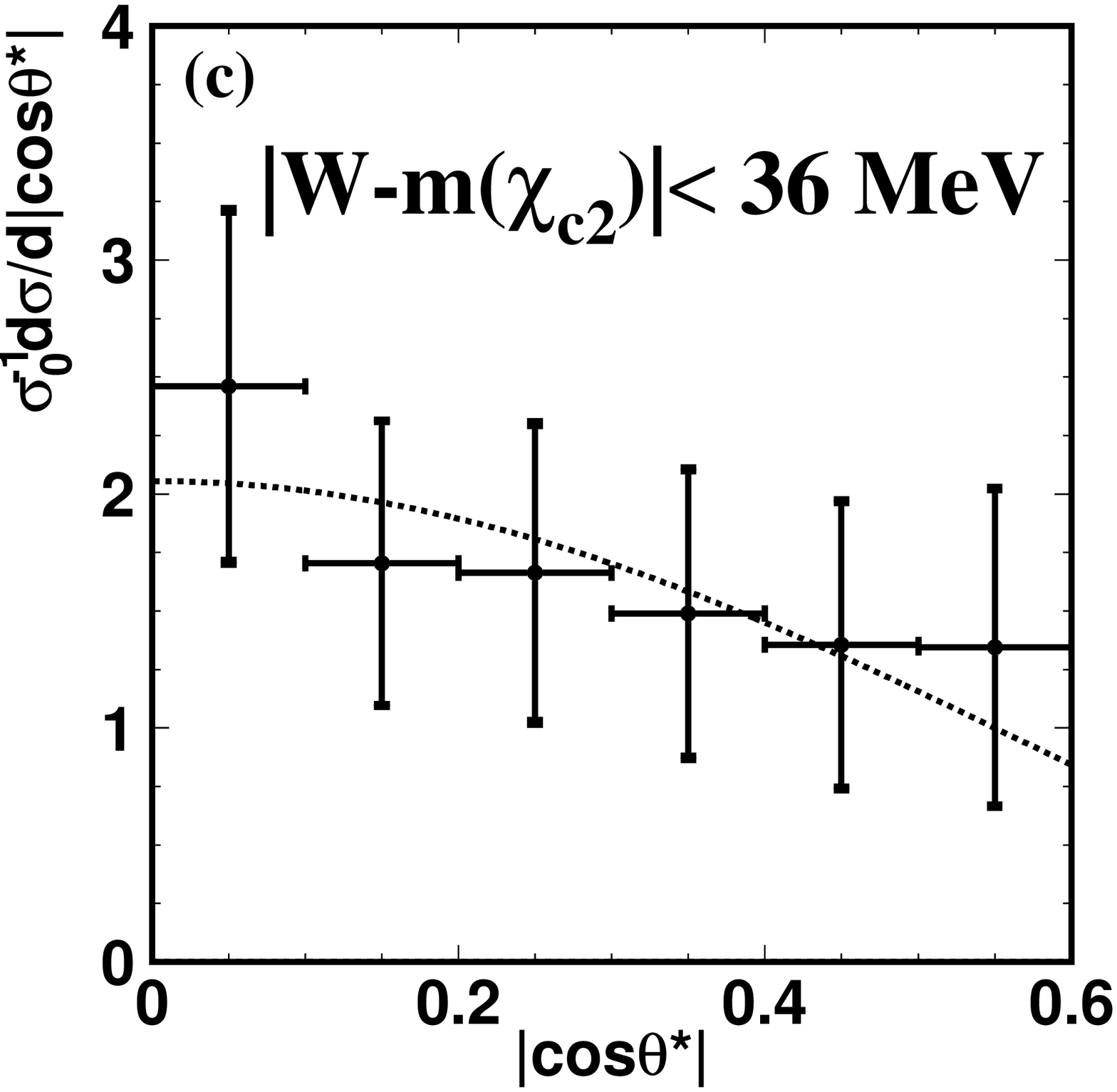}
\caption{(a1-a4) The angular distribution of the cross section, 
$\sigma^{-1}_0d\sigma/d|\cos\theta^*|$, in different $W$ ranges.
The solid curves are $1.227\sin^{-4}\theta^*$, which is the prediction 
of DKV. The dotted curves are the prediction of BC.
(b) The angular distribution in the $\chi_{c0}$ region; 
the dotted curve shows a flat distribution ($J$=0);
(c) The angular distribution in the $\chi_{c2}$ region;
the dotted curve shows the helicity 2 distribution 
($\propto\sin^4\theta^*$).
The errors indicated by short ticks are statistical only.}
\label{fig:diff}
\end{center}
\end{figure*}

The efficiencies $\epsilon(W, |\cos\theta^*|)$ are obtained from MC using 
the TREPS code \cite{TREPS} for $\ggr\ks\ks$ event generation. The TREPS 
code is also used for the luminosity function determination.
Trigger efficiencies are determined from the trigger simulator.
The typical values of the detection and trigger efficiency 
are 5-19\% and 90-95\%, respectively,
and grow with increasing $W$ and decreasing $|\cos\theta^*|$.
Differential cross sections normalized to the cross section 
integrated over the range 
$|\cos\theta^*|<0.6$ ($\sigma_0$) in different $W$ bins are shown 
in Fig.~\ref{fig:diff}(a). 
The angular distributions are consistent 
with both BC and DKV predictions up to $|\cos{\theta}^*|=0.5$.

\begin{figure}[htb]
\begin{center}
\includegraphics[width=0.45\textwidth]{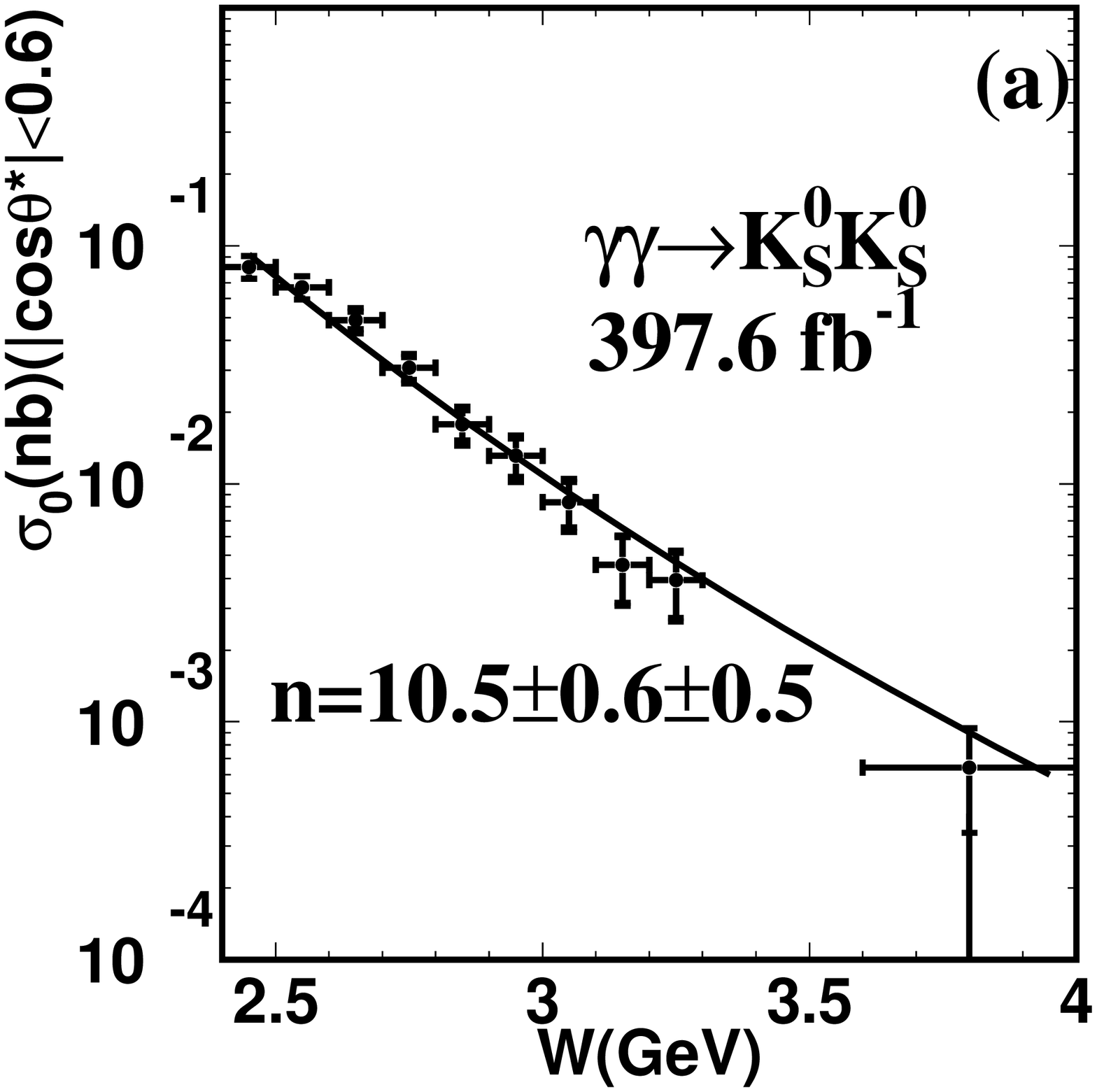}
\includegraphics[width=0.45\textwidth]{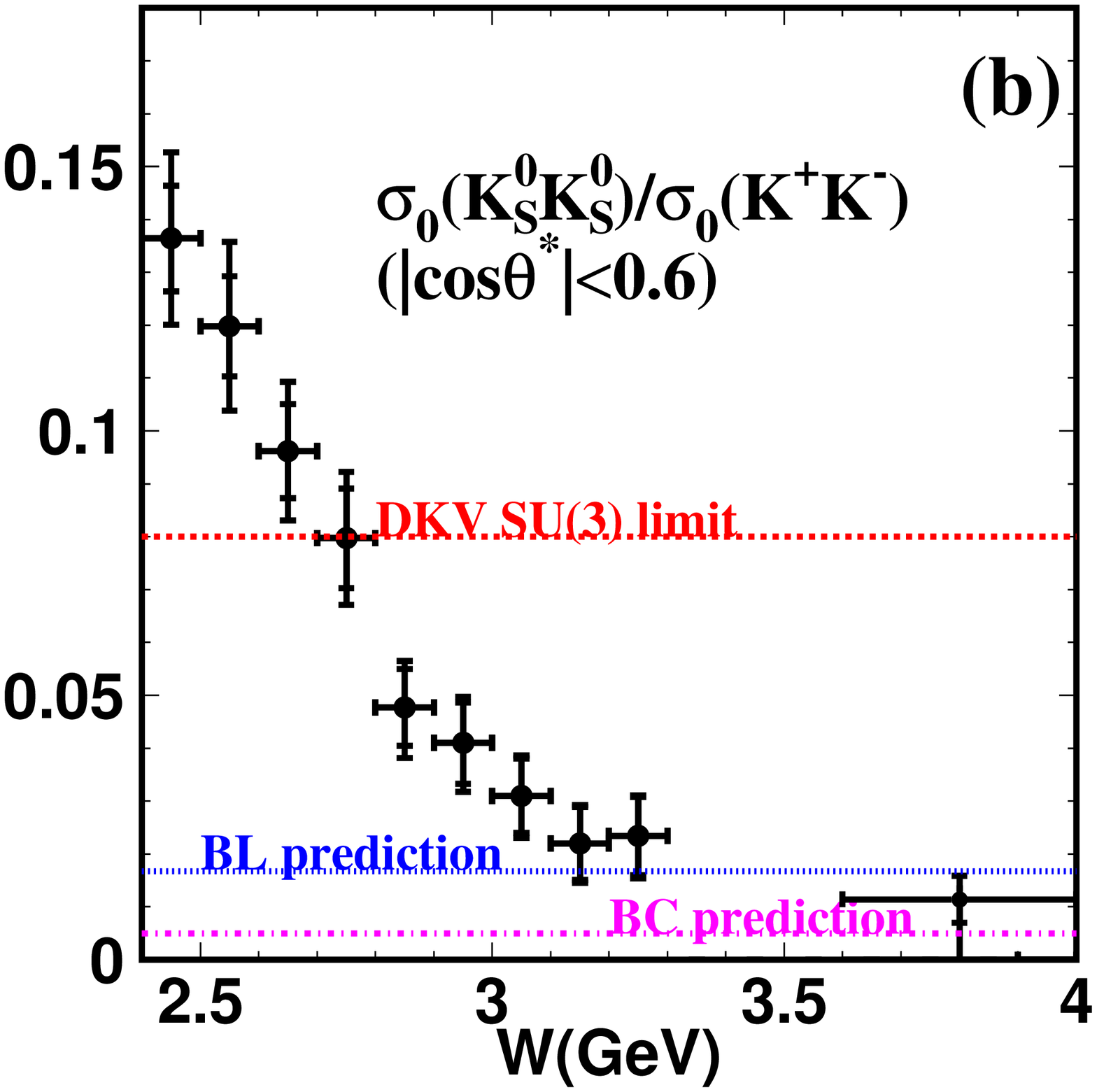}
\caption{(a) Total cross sections for $\ggr\ks\ks$ in the c.m. 
angular region $|\cos\theta^*|<0.6$. 
Here $n$ is the $W$-dependence ($\propto W^{-n}$). 
(b) The ratio $\sigma_0(\ks\ks)/\sigma_0(K^+K^-)$ versus 
$W$ in $|\cos\theta^*|<0.6$, 
where the $K^+K^-$ data are taken from the Belle measurement \cite{nkzw}.
The dotted line is the DKV prediction with the flavor symmetry assumption;
the dashed and dashed-dotted lines are the BL and BC predictions, 
respectively. 
The two sets of error bars show the
statistical and combined statistical+systematic errors, respectively.}
\label{fig:kkks}
\end{center}
\end{figure}

The angular distributions, $\sigma_0^{-1}d\sigma/d|\cos(\theta^*)|$, 
in the $\chi_{c0}$ and $\chi_{c2}$ regions
($|W-m(\chi_{c0})|<66$~MeV/$c^2$, $|W-m(\chi_{c2})|<36$~MeV/$c^2$) shown in 
Figs.~\ref{fig:diff}(b,c) are in good agreement with those expected for
the decays of the spin zero and  two particles.
The total cross section $\sigma_0$ as a function of $W$ is shown 
in Fig.~\ref{fig:kkks}(a) and listed in Table~\ref{tab:cs}.
The values of the total cross section for the range, $W$=3.3-3.6~$\GeV$, 
where the contribution from charmonium states is large, are omitted.

\begin{center}
\begin{threeparttable}
\begin{tabular}{c@{\hspace{0.3cm}}@{\hspace{0.3cm}}c@{\hspace{0.3cm}}
@{\hspace{0.3cm}}c}
\toprule[2pt]
{$W$($\GeV$)}& $N_{\rm ev}$ & {$\sigma_0$, nb} \\
\midrule[1pt]
{2.4-2.5}&  $226.3 \pm 15.4$& $0.0816 \pm 0.0056 \pm 0.0070$ \\
{2.5-2.6}&  $195.6 \pm 14.3$& $0.0671 \pm 0.0049 \pm 0.0057$ \\
{2.6-2.7}&  $137.9 \pm 12.0$& $0.0488 \pm 0.0042 \pm 0.0042$ \\
{2.7-2.8}&  $81.9 \pm 9.2$& $0.0307 \pm 0.0034 \pm 0.0027$ \\
{2.8-2.9}&  $46.8 \pm 6.9$& $0.0178 \pm 0.0026 ^{+0.0016}_{-0.0021}$ \\
{2.9-3.0}&  $31.1 \pm 5.7$& $0.0131 \pm 0.0024 ^{+0.0012}_{-0.0016}$ \\
{3.0-3.1}&  $21.4 \pm 4.7$& $0.0084 \pm 0.0018 ^{+0.0008}_{-0.0010}$ \\
{3.1-3.2}&  $10.7 \pm 3.3$& $0.0046 \pm 0.0014 ^{+0.0004}_{-0.0006}$ \\
{3.2-3.3}&  $10.7 \pm 3.3$& $0.0039 \pm 0.0012 ^{+0.0004}_{-0.0005}$ \\
{3.6-4.0}&  $5.0 \pm 2.2$ & $0.0006 \pm 0.0003 ^{+0.0001}_{-0.0006}$ ($< 0.0013$ at 90\%~CL)\\
\bottomrule[2pt]
\end{tabular}
\caption{Signal yields ($N_{\rm{ev}}$) and total cross 
sections ($\sigma_0$) for the process 
$\ggr\ks\ks$ in the angular range $|\cos\theta^*|<0.6$. The first 
and second errors are statistical and systematic, respectively.}
\label{tab:cs}
\end{threeparttable}
\end{center}
\section{Systematic errors}
The dominant systematic errors are summarized in Table~\ref{tab:sys}.
We assign 4\% to the uncertainty from trigger, 
which is determined by comparing the trigger efficiencies in the data sample and trigger simulation.
The uncertainty of $\ks$ reconstruction efficiency is estimated by
comparing the ratio of the number of $\ggr\ks\ks$ events with both $\ks$ mesons 
satisfying the selection requirements and that with only one $\ks$
satisfying the requirements in data and MC samples.
We take the efficiency difference between the data and MC $\ggr\ks\ks$ sample, which is 4.4\% for one $\ks$. 
The uncertainties in the background subtraction are estimated by 
fitting the background shape in the $|\Sigma\mathbf{p}^{ee}_t|$ distributions using second-order
polynomial functions and comparing the background fractions obtained
to those described above. 
The differences between the two calculations are taken as the corresponding 
systematic error in each energy range and are 
2.0\%, 2.0\%, and $^{+8.4}_{-2.6}$\% for $W$=2.4-2.6~GeV, 2.6-2.8~GeV, 
2.8-3.3~GeV, respectively.
For $W$=3.6-4.0~GeV, we conservatively assign the number of observed events as the systematic 
error in the background.
The 3.4-5.0\% systematic error for the luminosity function in 
the range $W$=2.4-4.0~GeV in Ref.~\cite{TREPS} is determined
from comparison of the kinematic distributions for the 
two-photon system in events generated with
TREPS to those from a QED calculation that includes
all order $\alpha^4$ diagrams \cite{AAFH}.
The total $W$-dependent systematic error is (8.5-12.1)$\%$.


\begin{center}
\begin{threeparttable}
\begin{tabular}{l@{\hspace{8mm}}c}
\toprule[2pt]
Source& Error, $\%$\\
\midrule[1pt]
Trigger efficiency & 4\\
Luminosity function& 3.4-5.0 \\
Background (for non-resonant analysis) & 2.0-8.4\\
$K^0_S$ reconstruction (per $\ks$)& 4.4\\
\vspace{0.5cm}
Integrated luminosity & 1.4\\
Total & 8.5-12.1\\
\bottomrule[2pt] 
\end{tabular}
\label{tab:sys}
\caption{Summary of systematic errors}
\end{threeparttable}
\end{center}

\section{Discussion}
The leading term in QCD calculations \cite{BL,BC} predicts a 
$\sim W^{-6}$ dependence of the cross sections $d\sigma/d\cos{\theta^*}
(\ggr M\overline{M})$.
However, the fit to the data in the range $W$=2.4-4.0~GeV gives 
a $W$-dependence ($\sigma_0\propto W^{-n}$) of $n=10.5\pm0.6\pm0.5$,
where the first error is statistical and the second is systematic. 
We conservatively estimate the systematic error on $n$ by artificially 
deforming the measured cross section values
assuming that the systematic errors are strongly correlated 
point-to-point, as in Ref.~\cite{nkzw}: 
we shift the $\sigma_0$ values at the two end bins by 
$\pm~1.5$ and $\mp~1.5$ times the systematic error, respectively, 
whereas each intermediate point is moved so that its shift follows a 
linear function of $W$ times its systematic error.
The average of the observed deviations in $n$ from its original value is 
taken as a final systematic error.
The value of $n$ indicates that, unlike $\gamma\gamma \to \pi^+\pi^-$ and 
$\gamma\gamma \to K^+K^-$~\cite{nkzw},
the current values of $W$ are not yet large enough to neglect power 
corrections in $\ggr\ks\ks$, 
which are not taken into account in the BL and BC predictions.



The ratio $\sigma_0(\ks\ks)/\sigma_0(K^+K^-)$ 
shown in Fig.~\ref{fig:kkks}(b)
decreases from $\sim$0.13 to $\sim$0.01 with increasing $W$.
This energy dependence is inconsistent with the DKV prediction 
that the ratio should be $\approx 2/25$ in the SU(3) symmetry limit.
Furthermore, it is difficult to explain the experimental result with 
the handbag model even if the effect of SU(3)-symmetry 
breaking is taken into account \cite{chernyak,gpd06}.
This indicates that the handbag model needs significant corrections. 

Since the experimental values of the ratio 
 $\sigma_0(\ks\ks)/\sigma_0(K^+K^-)$ approach the BL and BC
predictions at the highest measured energies $W \approx 4$~GeV, 
the leading term QCD calculations~\cite{BL,BC} may become
applicable for $\sigma(\ks\ks)$ at not much larger values of $W$.
\section{The two-photon decay width of $\chi_{cJ}$ resonances}
Measurements of $\ggr\ks\ks$ can also provide more precise results~\cite{PDG}
for the two-photon decay widths and branching fractions of 
the charmonium states since the continuum background is strongly suppressed.
By fitting the continuum $M(\ks\ks)$ distribution to an exponential distribution and 
parameterizing the charmonium peaks with a Breit-Wigner function for the 
$\chi_{c0}$ and Gaussian function for the narrow $\chi_{c2}$ 
with the masses and widths floating, 
$134\pm12$ $\chi_{c0}$ and $38\pm7$ $\chi_{c2}$ events are observed.
The masses and widths obtained from the fit taking into account
the detector resolution 
are consistent with the PDG values.
The $\chi_{c0}$($\chi_{c2}$) statistical significance is 
$22.7\sigma$ ($11.2\sigma$), where $\sigma$ is a standard deviation.
The statistical significance of the signals is obtained from the
$\sqrt{-2\ln(L_0/L_{\rm{max}})}$ values,
where $L_{0({\rm{max}})}$ is the likelihood without (with) the signal contribution,
with the joint estimation of the three parameters (mass, width, and yield are determined simultaneously). 
The two-photon decay width of the $\chi_{c0}$ or $\chi_{c2}$ 
can be obtained using the formula
\begin{eqnarray}
\Gamma_{\gamma\gamma}(\chi_{cJ})\times\mathcal{B}(\chi_{cJ}\rightarrow\ks\ks)=
\frac{Ym^2}{4(2J+1)\pi^2\mathcal{L}_{\gamma\gamma}(m)
\epsilon\mathcal{B}^2(K^0_S\rightarrow\pi^+\pi^-)\mathcal{L}_{\rm{int}}} ,
\end{eqnarray}
where $Y$ and $m$ are the yield and mass of the  $\chi_{cJ}$ charmonium
state, respectively.
The quantity $\epsilon$ denotes the product of the detector efficiency, 
trigger efficiency, and 
angular acceptance for the resonant decays.
In addition to the sources of systematic errors listed in 
Table~\ref{tab:sys},
the errors in the yield are 2.3$\%$ and 2.4$\%$ for the $\chi_{c0}$ and 
$\chi_{c2}$, respectively.
For $\chi_{c2}$ events we assume a pure helicity 2 state in MC
generation following the previous measurement~\cite{belpr} and
theoretical expectations~\cite{helth1,helth2}.
The directly measured values of the product 
$\Gamma_{\gamma\gamma}(\chi_{cJ})\mathcal{B}(\chi_{cJ}\rightarrow\ks\ks)$
are $7.00\pm0.65\pm0.6$~eV for the $\chi_{c0}$ and
$0.31\pm0.05\pm0.03$~eV for the $\chi_{c2}$.
Using the results of our previous measurement of $K^+K^-$ and
$\pi^+\pi^-$ production in $\gamma\gamma$ collisions~\cite{nkzw},
we determine the ratios 
$\mathcal{B}(\ks\ks)/\mathcal{B}(K^+K^-)$ 
and 
$\mathcal{B}(\ks\ks)/\mathcal{B}(\pi^+\pi^-)$ 
for the  $\chi_{c0}$ and $\chi_{c2}$, in which some common systematic
errors cancel. Here $\mathcal{B}(\pi^+\pi^-)$, $\mathcal{B}(K^+K^-)$ 
and $\mathcal{B}(\ks\ks)$ are the branching fractions for the $\chi_{cJ}$
decay to the corresponding final state. 
Using the world-average values of the branching fractions
${\mathcal{B}}(\chi_{c0} \to \ks\ks)=(2.8 \pm 0.7) \cdot 10^{-3}$ 
and ${\mathcal{B}}(\chi_{c2} \to \ks\ks)=(6.7 \pm 1.1) \cdot 10^{-4}$~\cite{PDG}, 
from the products of the widths and branching fractions given above 
we can extract the values of the two-photon
width that are shown in Table~\ref{tab:br}. 
The notation $br.$ indicates the systematic uncertainty from the 
branching fraction of $\chi_{cJ}\rightarrow\ks\ks$.
It can be seen that for both the $\chi_{c0}$ and $\chi_{c2}$ the value
of $\mathcal{B}(\ks\ks)/\mathcal{B}(K^+K^-)$ is compatible with 0.5 
as expected from isospin symmetry. 
The values of the two-photon widths of the $\chi_{c0(2)}$ charmonia
are consistent with those obtained from their total width and the
branching fractions for decay to two photons in Ref.~\cite{PDG}. 

\begin{center}
\begin{threeparttable}
\label{tab:br}
\begin{tabular}{c@{\hspace{2mm}}c@{\hspace{2mm}}c}
\toprule[2pt]
Resonance & $\chi_{c0}$ &$\chi_{c2}$ \\
\midrule[1pt]
$\Gamma_{\gamma\gamma}\mathcal{B}(\ks\ks)$, eV  &
$7.00\pm0.65\pm0.71$ &   $0.31\pm0.05\pm0.03$ \\
$\mathcal{B}(\ks\ks)/\mathcal{B}(K^+K^-)$ &
  $0.49\pm0.07\pm0.08$  &  $0.70\pm0.21\pm0.12$ \\
$\mathcal{B}(\ks\ks)/\mathcal{B}(\pi^+\pi^-)$ &
   $0.46\pm0.08\pm0.07$   &  $0.40\pm0.10\pm0.06$  \\ 
$\Gamma_{\gamma\gamma}$, keV  &
$2.50\pm0.23 \pm0.23 \pm0.62(br.)$ &   $0.46\pm0.08 \pm0.04 \pm0.08(br.)$ \\
\bottomrule[2pt]
\end{tabular}
\caption{The products of the two-photon width and the branching fraction,
ratios of the branching fractions,
and two-photon widths for the $\chi_{c0}$ and $\chi_{c2}$.
The notation $br.$ indicates the systematic uncertainty from the branching 
fraction of $\chi_{cJ}\rightarrow\ks\ks$}
\end{threeparttable}
\end{center}
\section{Conclusion}
Using a 397.6~fb$^{-1}$ data sample
accumulated with the Belle detector at KEKB,
the cross sections of the process $\ggr\ks\ks$ have been 
measured for the first time in the $W$ range from 2.4 GeV to 4.0 GeV 
with $|\cos\theta^*|<0.6$.
The overall $W$-dependent systematic uncertainty is 8.5-12.1$\%$.
The measured $W$-dependence ($\sigma_0\propto W^{-n}$) of $\ggr\ks\ks$ is 
$n=10.5\pm0.6\pm0.5$ from a fit to the data with $W$=2.4-4.0~$\GeV$,
indicating that, unlike $\gamma\gamma \to \pi^+\pi^-$ and $\gamma\gamma \to K^+K^-$, 
the $W$ values up to 3.3~GeV are not sufficiently
large to apply the leading term BL and BC predictions to $\ggr\ks\ks$.
The angular distribution in the range $|\cos\theta^*|<0.5$ is consistent with both BC and DKV.
The ratio $\sigma_0(\ggr\ks\ks)/\sigma_0(\ggr K^+K^-)$ 
decreases rapidly from $\sim$0.13 to $\sim$0.01 with increasing $W$ in contrast to the expectation from the DKV model.
Since the measured values of the cross section ratio approach the BL
and BC predictions in the highest energy bin, 3.6-4.0~GeV,
this may indicate that the leading term QCD calculations for $\sigma(\ggr \ks\ks)$ are already
applicable at $W$ values larger than $\sim 4$~GeV.
In addition, the products of the two-photon decay width and branching ratio 
to $\ks\ks$ for the $\chi_{c0}$ and $\chi_{c2}$
are found to be $7.00\pm0.65\pm0.71$~eV and $0.31\pm0.05\pm0.03$~eV.
\section{Acknowledgments}
We are grateful to S.~J.~Brodsky, V.L.~Chernyak, M.~Diehl, 
G.~Duplan\v{c}i\'{c}, P.~Kroll, H.-N.~Li, K.~Odagiri, and R.-C.~Verma 
for fruitful discussions.
We thank the KEKB group for the excellent operation of the
accelerator, the KEK cryogenics group for the efficient
operation of the solenoid, and the KEK computer group and
the National Institute of Informatics for valuable computing
and Super-SINET network support. We acknowledge support from
the Ministry of Education, Culture, Sports, Science, and
Technology of Japan and the Japan Society for the Promotion
of Science; the Australian Research Council and the
Australian Department of Education, Science and Training;
the National Science Foundation of China and the Knowledge
Innovation Program of the Chinese Academy of Sciences under
contract No.~10575109 and IHEP-U-503; the Department of
Science and Technology of India; 
the BK21 program of the Ministry of Education of Korea, 
the CHEP SRC program and Basic Research program 
(grant No.~R01-2005-000-10089-0) of the Korea Science and
Engineering Foundation, and the Pure Basic Research Group 
program of the Korea Research Foundation; 
the Polish State Committee for Scientific Research; 
the Ministry of Education and Science of the Russian
Federation and the Russian Federal Agency for Atomic Energy; 
the Slovenian Research Agency;  the Swiss
National Science Foundation; the National Science Council
(under Grant No. NSC 94-2112-M-008-027)
and the Ministry of Education of Taiwan; and the U.S.\
Department of Energy.

\end{document}